\documentclass[12pt]{article}
\usepackage{latexsym}
\usepackage[dvips]{graphicx}

\newcommand{\eq}{\begin{equation}}
\newcommand{\feq}{\end{equation}}
\newcommand{\eqn}{\begin{eqnarray}}
\newcommand{\feqn}{\end{eqnarray}}
\newcommand{\arr}{\begin{eqnarray*}}
\newcommand{\farr}{\end{eqnarray*}}
\newcommand{\beq}{\begin{equation}}
\newcommand{\eeq}{\end{equation}}
\newcommand{\bea}{\begin{eqnarray}}
\newcommand{\eea}{\end{eqnarray}}

\def\beq{\begin{equation}}
\def\eeq{\end{equation}}
\def\bea{\begin{eqnarray}}
\def\eea{\end{eqnarray}}
\def\bc{\begin{displaymath}}
\def\ec{\end{displaymath}}
\def\lb{\label}

\def\hatt{\hat\tau}

\def\la{\lambda}

\def\lb{\label}

\begin{document}
\begin{titlepage}
\vspace{.3cm}
\begin{center}
\renewcommand{\thefootnote}{\fnsymbol{footnote}}
{\Large \bf Horizons and the  Thermal Harmonic Oscillator }
\vfill
{\large \bf {M.~Cadoni\footnote{email: mariano.cadoni@ca.infn.it}}}\\
\renewcommand{\thefootnote}{\arabic{footnote}}
\setcounter{footnote}{0}
\vskip 7mm
{\small Dipartimento di Fisica,
Universit\`a di Cagliari, and INFN sezione di Cagliari, Cittadella
Universitaria 09042 Monserrato, ITALY}
 
\vspace*{0.4cm}
\end{center}
\vfill
\centerline{\bf Abstract}
\vskip 7mm
We show that two-dimensional anti-de Sitter spacetime (AdS$_{2}$) can be put in 
correspondence, holographically, both with the harmonic oscillator and 
the 
free particle. When  AdS$_{2}$ has a horizon the corresponding 
mechanical system is a thermal harmonic oscillator at temperature 
given by the Hawking temperature of  the horizon.  

\vfill
\end{titlepage}

The quantum description of black holes has represented, since the 
ground breaking work  of S. Hawking, a fundamental  challenge for 
theoretical physics. 
Classically,  a  black hole is  a  gravitating  system where information 
gets lost. At the semiclassical level the black hole emits particles 
with a thermal spectrum. This feature allows us to compute the 
 entropy of the black hole and to discover that it is
proportional to the total area of the black hole horizon.
The number of possible quantum states of a black hole should be therefore
explained by a field theory residing on the black hole horizon rather then in 
the volume. What we see here at work is the  ``Holographic 
Principle'',  stating that the total entropy of a system
localized in  a given region of space is bounded by the an expression 
proportional to the area of the boundary \cite{'tHooft:gx,Susskind:1994vu}.
Indications 
that the holographic picture  
could be a fundamental feature of the gravitational 
interaction come also from string theory and 
cosmology (For a recent review see \cite{Bousso:2002ju}). 

In particular,  string theory allows us in a number of cases to 
identify and count the quantum states of the black hole and to 
reproduce exactly the Bekenstein-Hawking result. Moreover,
 explicit realizations of the holographic principle has been 
found for anti-de Sitter (AdS) 
(and de Sitter) gravity, the so-called anti-de Sitter/conformal field 
theory (AdS/CFT) correspondence \cite{Maldacena:1997re,
Witten:1998qj,Gubser:1998bc}. 

Despite  of this considerable progress, the deep meaning of the 
holographic principle remains somehow mysterious.
Although everyone agrees that holography must be an essential feature 
of any theory of quantum gravity,  its relationship with the basic principles 
of  quantum mechanics and general relativity is poorly understood 
\cite{'tHooft:1999gk,'tHooft:2004ek}.
Explicit realizations of the holographic principle are known only 
in few cases. Moreover, they always take  the form of a duality,
in which the form of the mapping between bulk and boundary degrees of 
freedom is not explicitly known.

In this paper we discuss these issues in the context of 
two-dimensional (2D) AdS gravity. In this case the dual boundary 
theory is  De Alfaro-Fubini-Furlan (DFF) \cite{deAlfaro:1976je} conformal mechanics coupled 
with an external source  and the 
form of the
mapping between bulk and boundary degrees of freedom is explicitly 
known \cite{Cadoni:2000gm,Brigante:2002rv,Astorino:2002bj}. 
An other interesting feature of  the model is the fact that 
2D anti-de Sitter spacetime  (AdS$_{2}$) allows for three different 
parametrizations of the spacetime \cite{Cadoni:1993rn}. 
One of them 
exhibits  an event horizons, which is  analogous to the acceleration 
horizons of Rindler spacetime \cite{Cadoni:1994uf}. We can therefore use the mapping 
between the bulk and boundary degrees of freedom to put in 
correspondence the horizons (or more in general the 
spacetime structure) of the gravity theory  with the quantum 
mechanical (and thermal) description of the boundary mechanical system.
We will show  that  choosing appropriately the degrees of 
freedom,  the dual  boundary theory  becomes either an  harmonic oscillator or  
a free particle. In the parametrization where  AdS$_{2}$ has an horizon the corresponding 
mechanical system is a thermal harmonic oscillator with temperature 
given by the Hawking temperature of  the horizon.

 AdS$_{2}$ is a spacetime of constant 
negative curvature, $R=-2\lambda^{2}¥$.  It can be defined as an 
hyperboloid embedded in 3D Minkowski space \cite{Cadoni:1993rn}. Differently from higher 
dimensional cases,  AdS$_{2}$ admits three 
different parametrizations. Using  
Schwarzschild  coordinates the spacetime metric for  two of them can 
be written as \cite{Cadoni:1994uf}  
\beq\lb{me}
ds^2 = -\left(\la^{2}¥ r^{2} \pm a^{2}\right)dt^{2}+ \left(\la^{2}¥ r^{2}\pm 
a^{2}\right)^{-1}dr^{2}¥.
\eeq
The third parametrization is obtained setting $a=0$ in the previous 
equation. In the following we will denote these different 
parametrization of AdS$_{2}$ respectively as AdS$_{+}$, AdS$_{-}$ and
AdS$_{0}$ (Notice the change of notation with respect to Ref. \cite{ Cadoni:1994uf}).
The AdS$_{+}$ spacetime is full, geodetically complete, 2D AdS 
spacetime. It has cylindrical topology with two disconnected 
$r=\infty$ timelike, conformal boundaries, each of them having the topology of 
$S^{1}$.  Conversely,  the AdS$_{0}$ 
parametrization covers only part of the AdS hyperboloid. Only one of 
the two $r=\infty$ boundaries is visible and it has the topology of the 
line. The spacetime has an inner, null boundary at $r=0$.
Finally, the AdS$_{-}$ spacetime shares with AdS$_{0}$ the $r=\infty$ 
boundary structure but  has an event horizon at $r= a/\la$, whereas
$r=0$ becomes now spacelike.  

The three spacetimes are locally equivalent. The metrics in Eq. 
(\ref{me}) and that with $a=0$ can be transformed one into the other 
by means of a coordinate transformation \cite{ Cadoni:1994uf}. In this paper we 
will only need the asymptotic, $r=\infty$, form of these 
transformations,  i.e the transformation law for the time coordinates 
of the $r=\infty$ boundary of the AdS spacetime. Indicating with 
$\tau, t,{\hatt}$ the timelike coordinates of, respectively,  AdS$_{+}$, AdS$_{0}$ 
and  AdS$_{-}$, we have
\bea
\la t&=& \tan\frac{ a\la \tau}{2}, \quad  - \frac{\pi}{a\la}\le 
\tau\le \frac{\pi}{a\la}, \quad -\infty <t<\infty\lb{tt},\\
\la t&=& \pm \frac{1}{a}e^{a\la{\hat \tau}},\quad -\infty <\hat\tau<\infty\lb{tt1},
\eea
where the $\pm$ signs hold, respectively, for  $t>0$ and $t<0$.
Notice  that choosing the $+$ sign in Eq. (\ref{tt1}), the time 
coordinate ${\hatt}$ of the AdS$_{-}$ boundary covers only the region 
$t>0$ of the AdS$_{0}$ boundary.

The relationship between  AdS$_{-}$ and AdS$_{0}$ is analogue to that 
between Rindler and Minkowski spacetime. AdS$_{-}$ can be 
considered as the thermalization of AdS$_{0}$ at temperature, given by 
the Hawking temperature of the horizon, $T= 
a\la/2\pi$. Correspondingly, the vacuum for quantum fields in  AdS$_{0}$ 
will be detected by an  AdS$_{-}$ observer as a thermal flux of 
particles at temperature $T= 
a\la/2\pi$ \cite{Cadoni:1994uf} .

Introducing a scalar field $\Phi$ (the  dilaton) AdS$_{2}$ can be obtained as 
classical solution of  the 2D gravity 
action $S=\frac{1}{2}\int d^{2}x\sqrt{-g}\Phi (R+2\la^{2}¥)$. The  classical 
solution 
is now described by the metric (\ref{me}) endowed with a linear varying 
dilaton $\Phi=\la r$ (The most general solution contains a multiplicative 
integration constant $\Phi_{0}$, which is irrelevant for our 
purposes and has been  set equal to 1) . The presence of the dilaton is crucial.
It enable us to interpret AdS$_{-}$  as a 2D black hole. In fact 
$\Phi^{-1}¥$ is proportional to the (coordinate dependent) 2D newton 
constant,
so that $r=0$ can be considered as a singularity. The mass, temperature 
and entropy of the black hole are given by 
\beq\lb{mts}
M=\frac{1}{2}\la a^{2}, \quad T=\frac{a\la}{2\pi},\quad S= 
2¥\pi a.
\feq
In this context the AdS$_{0}$ spacetime has to be considered as the
$M=T=0$ solution, whereas AdS$_{+}$ describes  a ``naked'' singularity,
a black hole with negative mass $M=-(1/2)\la a^{2}$.

Two-dimensional AdS gravity induces on the spacetime boundary a 
conformal invariant dynamics \cite{Cadoni:2000gm}. The boundary theory has the form of 
de Alfaro-Fubini-Furlan (DFF) conformal mechanics coupled with an 
external source. Moreover, the thermodynamical entropy (\ref{mts}) can 
be exactly reproduced by counting states in the boundary conformal 
theory \cite{Cadoni:1998sg,Cadoni:1999ja,Navarro:1999qy,
Catelani:2000gn,Cadoni:2000fq,Jing:2000yn,
Cadoni:2000ah,Carlip:2002be,Silva:2002jq,
Medved:2002dw,Kang:2004js,Fursaev:2004qz}. The  equations describing the boundary dynamics can be 
derived considering the asymptotic symmetry group of  AdS$_{2}$ and 
the related large $r$ behavior of the metric and of the dilaton 
\beq\lb{bc}
g_{tt}\sim -\la^{2} r^{2}+ \gamma_{tt}(t),\quad g_{tt}\sim {1\over \la^{2} r^{2}}+
{\gamma_{rr}(t)\over \la^{4}r^{4}},\quad \Phi\sim\la \rho(t) r+
{\gamma_{\Phi}(t)\over 2\la r}.
\eeq
The fields $\gamma_{tt},\gamma_{rr}, \gamma_{\Phi},\rho$ represent  boundary and dilaton  
deformations.
The   field equations for the 2D metric and dilaton projected on 
the boundary produce the dynamical equations \cite{Cadoni:2000gm}
\bea
\la^{-2}¥\ddot \rho-\rho \gamma+\beta&=&0,\lb{be}\\
\dot \rho \gamma +\dot \beta&=&0, \lb{be1}
\eea
where $\gamma=\gamma_{tt}-{(1/2)}\gamma_{rr}$,  $\beta={(1/ 
2)}\rho\gamma_{rr}+\gamma_{\Phi}$ and the dot denotes derivation with 
respect to time.

Two-dimensional dilaton gravity has no physical  propagating,    bulk degrees of 
freedom. However, Eqs. (\ref{be}), (\ref{be1}) tell us that on the  boundary  of 
the AdS spacetime 
there are dynamical degrees of freedom. This phenomenon has been 
already observed  in an other topological theory, namely 3D 
gravity:
pure gauge  bulk degrees of freedom become physical on the boundary 
\cite{Carlip:1996yb}. 
In the case under consideration the only  dynamical boundary degree of 
freedom is the dilaton deformation $\rho$. The other deformations 
$\gamma, \beta$ appearing in Eqs. (\ref{be}), (\ref{be1})  are not dynamical and 
are related to the diffeomorphism invariance of the 2D bulk theory. 
Under  the action of infinitesimal 
diffeomorphisms of the bulk that  leave the form (\ref{bc}) of the 
metric invariant,  $\rho, \gamma, \beta$ transform as conformal fields  
of weights $-1,2,1$ respectively \cite{Cadoni:2000gm}. 

These bulk 
transformations are realized on the AdS boundary as the $diff_{1}$ 
group  of  time reparametrizations, whose generators satisfy a 
Virasoro algebra \cite{Cadoni:1998sg, Cadoni:1999ja}.  
In principle one could then set $\gamma$ and/or 
$\beta$ to a constant just by using the diffeomorphism invariance of
the bulk gravity theory. This is not possible if we want to preserve 
full conformal invariance of the  boundary theory.
Conformal invariance is crucial if one 
wants to use the boundary theory to give a microscopical 
interpretation of the thermodynamical entropy of the 2D black hole 
\cite{Cadoni:1998sg,Cadoni:1999ja}. In this case $\gamma$ 
and $\beta$ have to be considered as external sources that encode the 
information about the diffeomorphism invariance of the 2D gravity theory. 
This leads to the interpretation of the dynamical system (\ref{be}), (\ref{be1}) as 
DFF conformal mechanics coupled to an external source \cite{Cadoni:2000gm}.
 
In this paper we will not require invariance of the boundary theory 
under the  $diff_{1}$ conformal group, so that we can hold   $\gamma$ 
constant.  
For $\gamma= const.$  Eq.  (\ref{be1}) can 
be easily integrated to give  $\beta= -\rho\gamma +C$ where $C$ is an 
integration constant. Setting $C=0$ and using this equation into Eq. (\ref{be}) and 
one gets
\beq\lb{ao}
\ddot \rho-2\la^{2}¥ \rho \gamma =0,\lb{bbe1}.\\
\eeq
This equation describes an elementary mechanical system.
It is the equation of motion coming  from the lagrangian 
\beq\lb{la}
{\cal {L}}= \frac{{\dot q}^{2}}{2}-\frac{\omega^{2}}{2} q^{2},
\feq
with 
\beq\lb{f1}
q=\frac{\rho}{\sqrt\la}, \quad \omega^{2}= -2 \la^{2}\gamma.
\feq
Depending on the sign of $\gamma$,  the Lagrangian (\ref{la}) 
describes a harmonic oscillator ($\gamma<0$) a free particle 
($\gamma=0$) and a harmonic oscillator with imaginary frequency 
($\gamma>0$). 

We can now easily identify the mechanical system associated with 
the three AdS spacetimes discussed in the previous section. AdS$_{+}$ 
has $\gamma=- a^{2}/2=-M/\la$, its counterpart on the boundary is a 
harmonic oscillator with frequency 
\beq\lb{fre}
\omega= a \la= \sqrt{2M\la}.
\feq
AdS$_{0}$ is characterized by $\gamma=0$, it corresponds to a free 
particle. AdS$_{-}$, the black hole, has $\gamma= 
a^{2}/2=M/\la$, corresponding to a harmonic oscillator with  
imaginary frequency $-i\omega$, with $\omega$ given by Eq. (\ref{fre}).

In establishing the correspondence between the 2D metric 
(\ref{me}) and the mechanical system (\ref{la}), we have completely 
forgotten the global features of the boundary  of the   
 AdS spacetime.  The time coordinate of the mechanical system takes 
 its value on the timelike boundary of AdS$_{2}$. 
The time coordinates of the  systems with  $\gamma<0$,
$\gamma=0$ and $\gamma>0$ are given by $\tau, t,$ and ${\hatt}$ 
respectively. They are related one with the other by  
the  
 transformations (\ref{tt}), (\ref{tt1}).
For this reason we need a    general formalism, 
describing    the time  evolution of a mechanical system, which 
allows for time reparametrizations such as those given in Eq. 
(\ref{tt}), (\ref{tt1}).
A general formalism with these features has been 
proposed by DFF in their investigations of conformal mechanics 
\cite{deAlfaro:1976je}. 

The  action 
\beq\lb{fp}
A= \frac{1}{2}\int{\dot Q^{2}} dt,
\eeq
describes a free particle 
and is invariant under the ``little'' conformal group generated by
translations $H$, dilatations $D$ and special conformal 
transformations $K$. 
The most general conformal invariant model is  given by
the DFF Lagrangian ${\cal {L}}= (1/2)({\dot Q^{2}}- g/Q^{2})$. However,  in this 
paper we only need to consider the particular case $g=0$.
The generators obey the algebra  $[H,D]=iH,\,[K,D]=-iK,\,[H,K]=2iD$.
It has been noticed \cite{deAlfaro:1976je} that any linear combination of the three 
generators
\beq\lb{lc}
G=uH+vD+wK,
\eeq
is a constant of motion, therefore can be used to generate the 
dynamics  of the system. The generators $G$ have been classified by 
DFF in three classes depending on the sign of the determinant 
$\Delta=v^{2}-4uw$. For $\Delta<0$ $G$ is a compact operator,  its 
spectrum is discrete and bounded from below, whereas its eigenstates 
are normalizable. $\Delta>0$ corresponds 
to a non-compact  $G$, whose spectrum is unbounded from below. 
Finally, $\Delta=0$ corresponds to ``parabolic'' generators, whose 
spectrum is continuos and bounded from below. 

DFF argued that only 
operators with $\Delta<0$ lead to time evolution laws that are 
physically acceptable. Moreover, they can be used to solve the 
well-known infrared problem, which appears when time evolution is 
generated by parabolic generators (like $H$).
We will show later on this paper that in our framework we can also give a physical 
meaning to generators with $\Delta>0$.

The operator $G$ generates time evolution of the system but in terms 
of a new time variable $\tau$ (and  a new field $q(\tau)$) given by
\beq\lb{nt}
d\tau= \frac{dt}{u+vt+wt^{2}},\quad q(\tau)= 
\frac{Q(t)}{\left(u+vt+wt^{2}\right)^{1/2}¥}.
\eeq
The action (\ref{fp}), expressed in terms of the new variables takes the form \cite{deAlfaro:1976je}
\beq\lb{fp1}
A=  \frac{1}{2}\int d\tau\left({\dot q^{2}} + \frac{\Delta}{4} q^{2}\right).
\eeq
Choosing appropriately the parameters $u,v,w$ in Eq. (\ref{lc}), we 
can obtain the lagrangian (\ref{la}) from Eq. (\ref{fp1}).

For $v=w=0$ and $u=1$, time evolution (with respect to the time 
coordinate $-\infty <t<\infty $) is generated by the operator 
$G=H$. $\Delta=0$ and Eq. (\ref{fp1}) describes the free particle of 
Eq. (\ref{la}) (with $\omega=0$) associated with the AdS$_{0}$ spacetime.

For $u=a, v=0$ and $w=a\la^{2}$ the generator $G$ is compact  
($\Delta=-4 a^{2}\la^{2}$) and given by 
\beq\lb{te}
G=\la a(\la K+ \la^{-1}H).
\feq
The operator $G$ generates  time evolution with respect to 
the time 
coordinate $\tau$ of Eq. (\ref{tt}).
The action (\ref{fp1}) describes the harmonic oscillator of 
Eq. (\ref{la}), with $\omega$ given by Eq. (\ref{fre}), i.e the 
mechanical system associated with the AdS$_{+}$ spacetime. 
The transformation between the time coordinates $t$ 
and $\tau$ can be obtained integrating the first Eq. (\ref {nt}). One 
finds  that after the  rescaling $\tau\to\tau/2$  this transformation 
matches exactly the transformation (\ref{tt}) ( The rescaling 
is necessary owing to the different conformal weights of the  
fields $q$ in Eq. (\ref{la}) and (\ref{fp}), which are respectively 
$-1$ and $-1/2$). 

Quantizing the classical system we get the usual quantum harmonic 
oscillator. In our context the spectral properties of the 
quantum harmonic oscillator are a simple consequence of the 
compactness of the  time-evolution operator $G$ and of the 
periodicity  of the time coordinate $\tau$ 

For  $u=w=0$ and $v=2a\la$,  $\Delta= 4a^{2}\la^{2}$ is 
positive and the  operator
\beq\lb{te1}
G= 2a\la D
\feq
is noncompact and generates time evolution with respect to the time 
$\hatt$ of Eq. (\ref{tt1}). 
The action 
(\ref{fp1}) describes the harmonic oscillator with imaginary 
frequency  $\omega=-i a\la$  of
Eq. (\ref{la}), associated with  AdS$_{-}$.
A DFF model with a harmonic potential having the wrong sign has been 
also found considering the motion of a charged particle near the 
horizon of an extreme Reissner-Nordstr\"{o}m solution \cite{Mignemi:2001uz}.
Some related aspects of the conformal symmetry for particles moving in 
a AdS background have been also discussed in Ref. \cite{Moretti:2002mp}

Integrating Eq. (\ref{nt}) and rescaling the time 
coordinate $\tau\to\hatt/2$  one gets the 
transformation (\ref{tt1}) relating the time coordinates on the 
boundary of
AdS$_{-}$ and AdS$_{0}$. The coordinate transformation (\ref{tt1}) 
exchanges the generator of time translations $H$ with that of 
dilatations $D$. For this reason it can be considered as the 
one-dimensional analogue of the plane-cylinder transformation for 
2D conformal field theories.

Differently from AdS$_{+}$,   the AdS$_{-}$ spacetime presents an event 
horizon.  The presence of a spacetime region that is causally 
disconnected form the outside has  a strong impact on the features of 
the associated  mechanical system defined on the spacetime boundary. 
A quantum field theory in the AdS$_{-}$ background will generally have 
problems, owing to the necessity of tracing out 
the degrees of freedom behind the horizon.  This is a well-known 
effect that leads to the interpretation of  AdS$_{-}$ as the 
thermalization of AdS$_{0}$ \cite{Cadoni:1994uf}. We therefore expect that also the 
boundary mechanical system will be plagued   by the same problem.  
It shows up in two related issues. First, the 
Lagrangian (\ref{la}) in this case has a potential 
$V(q)= -\frac{1}{2} \omega^{2}q^{2}$, which is unbounded from below.
Correspondingly, the operator $G$ of Eq. (\ref{te1}),  which generates 
the time evolution  with respect to the time coordinate $\hat \tau$, 
is non-compact. The system does not seem to have a physically 
reasonable spectrum. 
Second, an observer using the  time coordinate $\hat \tau$ cannot 
see  
the whole history of the system as seen by an observer using  the time coordinate 
$t$ (or $\tau$) because, owing to Eq (\ref{tt1}),  for $-\infty<\hat 
\tau<\infty$,  $0<t<\infty$. The operator $G$ cannot generate unitary 
time evolution over the full range of the time variable $t$.
These troubles led DFF to discard the class of generators with $\Delta>0$, 
as physically unacceptable.

In our context the above mentioned problems have a natural 
interpretation: they are the boundary counterpart of the presence of 
an horizon on the 2D bulk. We can therefore hope to give to the 
mechanical system (\ref{la}) with $\omega^{2}¥<0$ a reasonable 
physical meaning. 
We will show  using two different methods that the system can be 
interpreted as a {\sl thermal quantum harmonic oscillator at 
 the horizon temperature} (\ref{mts}).

Formally, the mechanical system under consideration  is the analytic 
continuation $\omega \to - i \omega $ of the usual harmonic oscillator.
Let us now consider the time evolution of an energy 
eigenstate $|u_{n}\rangle$, 
of  the harmonic oscillator with eigenvalue $E_{n}$,
\beq \lb{ff}
|u_{n}(\tau)\rangle= e^{-iE_{n}\tau}|u_{n}(0)\rangle.
\feq
Because for the harmonic oscillator $E_{n}$ depends linearly on 
$\omega$ by analytical continuation, $\omega\to -i \omega$, we get
$|u_{n}(\tau)\rangle= e^{-E_{n}\tau}|u_{n}(0)\rangle$.  After a period 
$2\pi/\omega$ we get the (unnormalized) probability distribution
\beq\lb{f2}
\rho_{n}= \langle u_{n}(0)|u_{n}(\frac{2\pi}{\omega})\rangle= e^{-\beta E_{n}},
\feq
where $\beta= 2\pi/\omega= 2\pi/a\lambda$. 
Eq. (\ref{f2}) describes a thermal distribution of harmonic 
oscillators at temperature $T= a\la/2\pi$, the  
temperature of the horizon.

The same result can be obtained using a method which is analog to 
that used in quantizing free fields in the AdS$_{-}$ and AdS$_{0}$
background \cite{Cadoni:1994uf}. The key point is that the time coordinate $\hat 
\tau$ covers only  the region  $0\le t<\infty$.
The time evolution of a generic quantum state with respect to the 
time $\hat\tau$ is given by 
\beq\lb{te3}
| \psi_{I}¥(\hat \tau)\rangle=\sum_{n}c_{n} 
e^{-iE_{n}(\hatt-\hatt_{0})}|u_{n}¥(\hatt_{0})\rangle, \quad  a\la t= \exp{a\la 
\hat\tau} .
\feq
This time evolution law holds only in the region $I: t,t_{0}\ge 0$, 
where $t,t_{0}$ are the images of $\hatt,\hatt_{0}$ trough the 
transformation (\ref{tt1}) taken with the plus sign. 
We can continue  Eq. (\ref{te3}) into the region $II: t,t_{0}\le 0$
using Eq. (\ref{tt1}) with the minus sign and writing,
\beq\lb{te4}
| \psi_{II}¥(\hat \tau)\rangle=\sum_{n}c_{n} 
e^{-iE_{n}(\hatt-\hatt_{0})}|u_{n}¥(\hatt_{0})\rangle, \quad  a\la t= -\exp{a\la \hatt} .
\feq
The problem is that owing to the change of sign at $t=0$ , 
$e^{-iE_{n}(\hatt-\hatt_{0})}$ is not analytic at that point. As a 
consequence  time evolution will be unitary separately on the 
regions $I$ and $II$ but will become non-unitary when we cross $t=0$.
We can recover   analyticity at $t=0$ by 
defining 
 \beq\lb{te2}
| \psi(\hat \tau)\rangle=A\sum_{n}c_{n} 
e^{-\frac{\pi E_{n}}{a\la}¥}¥e^{-iE_{n}(\hatt-\hatt_{0})}|u_{n}¥(\hatt_{0})\rangle, 
\feq
where $A$  is a normalization constant and now $\hatt$ and $\hatt_{0}$ belong 
respectively to the regions $I$ and $II$. Time evolution is now not 
unitary but   
analytic at $t=0$ because the time evolution factor in Eq. 
(\ref{te2}) is proportional to $(t)^{-i E_{n}/a\la}$.
If the system is in an  eigenstate $|u_{n}¥\rangle$ at the time 
$\hatt_{0}$, the probability of finding it in the same state at the time 
$\hatt$ will be given by 
\beq\lb{f4}
P_{n}= |A|^{2} e^{-\frac{2\pi E_{n}}{a\la}
}
.
\feq
The normalization constant can be given in terms of the partition 
function $Z$, so that we finally find
\beq\lb{f5}
P_{n}=  \frac{e^{- \beta E_{n}}}{Z},
\feq
i.e a thermal distribution at temperature  given by the Hawking 
temperature of the horizon, $T=a\la/2\pi$.

Our simple example tells us that the holographic 
principle, quantum mechanics and the coarse graining of information 
of the thermal  description are intimately intertwined with the 
presence  of an horizon and with the topology of the spacetime.  
On the one hand our   results give support  to the usual thermal  interpretation of  
horizons. Constructing an holographic dual of the horizon in terms of 
an elementary mechanical system, we have found it still has the features of 
a thermal ensemble: a thermal ensemble of harmonic oscillators.

\end{document}